\documentclass[pre,showpacs,twocolumn,amsfonts,amsmath,nofootinbib]{revtex4}%
 \usepackage{pslatex}
 \usepackage{graphicx}% Include figure files

\begin{document}

 \title{Deterministic Modularity Optimization}
  \date{\today}
   \author{Sune Lehmann}
   \email{slj@imm.dtu.dk}
   \author{Lars Kai Hansen}
   \affiliation{Informatics and Mathematical Modeling, Technical
    University of Denmark, Building 321, DK-2800 Kgs.~Lyngby, Denmark.}

\newcommand{\note}[1]{\textsf{\textbf{(}#1\textbf{)}}}
\newcommand{\tr}{\mathrm{Tr}}
\newcommand{\sbxs}[1]{\framebox[1cm][c]{%
\begin{minipage}[c]{1cm} \begin{center}\vspace{.25cm}  #1 \vspace{.25cm} \end{center}\end{minipage}%
}}

\begin{abstract}
We study community structure of networks. We have developed a scheme
for maximizing the modularity $Q$ \cite{newman:04a} based on mean
field methods. Further, we have defined a simple family of random
networks with community structure; we understand the behavior of
these networks analytically. Using these networks, we show how the
mean field methods display better performance than previously known
deterministic methods for optimization of $Q$.
\end{abstract}


\maketitle

\section{Introduction}\label{sec:introduction}
A theoretical foundation for understanding complex networks has
developed rapidly over the course of the past few years
\cite{albert:02,dorogovtsev:02,newman:03a}. More recently, the
subject of detecting network \emph{communities} has gained an large
amount of attention, for reviews see
Refs~\cite{newman:04d,danon:05a}. Community structure describes the
property of many networks that nodes divide into modules with dense
connections between the members of each module and sparser
connections between modules.

In spite of a tremendous research effort, the mathematical tools
developed to describe the structure of large complex networks are
continuously being refined and redefined. Essential features related
to network structure and topology are not necessarily captured by
traditional global features such as the average degree, degree
distribution, average path length, clustering coefficient, etc. In
order to understand complex networks, we need to develop new
measures that capture these structural properties. Understanding
community structures is an important step towards developing a range
of tools that can provide a deeper and more systematic understanding
of complex networks. One important reason is that modules in
networks can show quite heterogenic behavior \cite{newman:06b}, that
is, the link structure of modules can vary significantly from module
to module. For such heterogenic systems, global measures can be
directly misleading. Also, in practical applications of network
theory, knowledge of the community structure of a given network is
important. Access to the modular structure of the internet could
help search engines supply more relevant responses to queries on
terms that belong to several distinct communities\footnote{Some
search engines have begun implementing related ideas, see for
example \emph{Clusty, the Clustering Engine} (http://clusty.com/).
There is, however, still considerable room for improvement.}. In
biological networks, modules can correspond to functional units of
some biological system \cite{barabasi:04}.

\section{The Modularity}\label{sec:previouswork}
This section is devoted to an analysis of the modularity $Q$.
Identifying communities in a graph has a long history in mathematics
and computer science \cite{chung:97,newman:04d}. One obvious way to
partition a graph into $C$ communities is distribute nodes into the
communities, such that the number of links connecting the different
modules of the network is minimized. The minimal number of
connecting links is called the \emph{cut size} $R$ of the network.

Consider an unweighted and undirected graph with $n$ nodes and $m$
links. This network can be represented by an adjacency matrix
$\mathbf{A}$ with elements
\begin{equation}\label{eq:adjacency}
    A_{ij} = \left\{
               \begin{array}{ll}
                 1, & \hbox{if there is a link joining nodes $i$ and $j$;} \\
                 0 & \hbox{otherwise.}
               \end{array}
             \right.
\end{equation}
This matrix is symmetric with $2m$ entries. The degree $k_i$ of node
$i$ is given by $k_i =\sum_{j} A_{ij}$. Let us express the cut-size
in terms of $\mathbf{A}$; we find that
\begin{equation}\label{eq:cutsize}
    R = \frac{1}{2}\sum_{i,j} A_{ij}[1-\delta(c_i,c_j)],
\end{equation}
where $c_i$ is the community to which node $i$ belongs and
$\delta(\alpha,\beta)=1$ if $\alpha=\beta$ and
$\delta(\alpha,\beta)=0$ if $\alpha\ne\beta$. Minimizing $R$ is an
integer programming problem that can be solved exactly in polynomial
time \cite{goldschmidt:88}. The leading order of the polynomial,
however, is $n^{C^2}$ which very expensive for even very small
networks. Due to this fact, most graph partitioning has been based
on spectral methods (more below).

Newman has argued \cite{newman:04d,newman:06a,newman:06b} that $R$
is not the right quantity to minimize in the context of complex
networks. There are several reasons for this: First of all, the
notion of cut-size does not capture the essence of our `definition'
of network as a tendency for nodes to divide into modules with dense
connections between the members of module and sparser connections
between modules. According to Newman, a good division is not
necessarily one, in which there are few edges between the modules,
it is one where there are fewer edges than expected. There are other
problems with $R$: If we set the community sizes free, minimizing
$R$ will tend to favor small communities, thus the use of $R$ forces
us to decide on and set the sizes of the communities in advance.

As a solution to these problems, Girvan and Newman propose the
modularity $Q$ of a network \cite{newman:04a}, defined as
\begin{equation}\label{eq:modularity}
    Q = \frac{1}{2m}\sum_{ij} [A_{ij} - P_{ij}]\delta(c_i,c_j).
\end{equation}
The $P_{ij}$, here, are a null model, designed to encapsulate the
`more edges than expected' part of the intuitive network definition.
It denotes the probability that a link exists between node $i$ and
$j$. Thus, if we know nothing about the graph, an obvious choice
would be to set $P_{ij} = p$, where $p$ is some constant
probability. However, we know that the degree distributions of real
networks are often far from random, therefore the choice of $P_{ij}
\sim k_ik_j$ is sensible; this model implies that the probability of
a link existing between two nodes is proportional to the degree of
the two nodes in question. We will make exclusive use of this null
model in the following; the properly normalized version is $P_{ij} =
(k_i k_j)/(2m)$. It is axiomatically demanded that that $Q = 0$ when
all nodes are placed in one single community. This constrains the
$P_{ij}$ such that
\begin{equation}\label{eq:constraintQ}
    \sum_{ij} P_{ij} = 2m,
\end{equation}
we also note that $\mathbf{P} = (\mathbf{P})^T$, which follows from
the symmetry of $\mathbf{A}$.

Comparing Eqs.~\eqref{eq:cutsize} and~\eqref{eq:modularity}, we
notice that there are two differences between $Q$ and $R$. The first
is that $Q$ implies that we \emph{maximize} the number of
intra-community links instead of minimizing the the number of
inter-community links as is the case for $R$---this is the
difference between multiplying by $\delta(c_i,c_j)$ and
$[1-\delta(c_i,c_j)]$. The second difference lies in the the
introduction of the $P_{ij}$ in Equation~\eqref{eq:modularity}. The
subtraction of $P_{ij}$ serves to incorporate information about the
inter-community links into the quantity we are optimizing.

Use of modularity to identify network communities is not, however,
completely unproblematic. Criticism has been raised by Fortunato and
Barth\'elemy \cite{fortunato:06} who point out that the $Q$ measure
has a resolution limit. This stems from the fact that the null model
$P_{ij}\sim k_i k_j$ can be misleading. In a large network, the
expected number of links between two small modules is small and
thus, a single link between two such modules is enough to join them
into a single community. A variation of the same criticism has been
raised by Rosvall and Bergstrom~\cite{rosvall:06a}. These authors
point out that the normalization of $P_{ij}$ by the total number of
links $m$ has the effect that if one adds a distinct (not connected
to the remaining network) module to the network being analyzed and
partition the whole network again allowing for an additional module,
the division of the original modules can shift substantially due to
the increase of $m$.

In spite of these problems, the modularity is a highly interesting
method for detecting communities in complex networks when we keep in
mind the limitations pointed out above. What makes the modularity
particularly interesting compared to other clustering methods is its
ability to inform us of the optimal number of communities for a
given network\footnote{This ability to estimate the number of
communities, however, stems from the introduction of the $P_{ij}$
term in the Eq.~\eqref{eq:modularity} and is therefore directly
linked to the conceptual problems with $Q$ mentioned in the previous
paragraph.}.

\section{Spectral Optimization of Modularity}
The question of finding the optimal $Q$ is a discrete optimization
problem. We can estimate the size of the space we must search to
find the maximum. The number of ways to divide $n$ vertices into $C$
non-empty sets (communities) is given by the Stirling number of the
second kind $S_n^{(C)}$ \cite{mathworld}. Since we do not know the
number of communities that will maximize $Q$ before we begin
dividing the network, we need to examine a total of $\sum_{C=2}^n
S_n^{(C)}$ community divisions \cite{newman:04b}. Even for small
networks, this is an enormous space, which renders exhaustive search
out of the question.

Motivated by the success of spectral methods in graph partitioning,
Newman suggests a spectral optimization of $Q$ \cite{newman:06a}. We
define a matrix, called the modularity matrix $\mathbf{B} =
\mathbf{A}-\mathbf{P}$ and an $(n \times C)$ \emph{community matrix}
$\mathbf{S}$. Each column of $\mathbf{S}$ corresponds to a community
of the graph and each row corresponds to a node, such that the
elements
\begin{equation}\label{eq:Sdef}
    S_{ic} = \left\{
               \begin{array}{ll}
                 1, & \hbox{if node $i$ belongs to community $c$;} \\
                 0 & \hbox{otherwise.}
               \end{array}
             \right.
\end{equation}
Since each node can only belong to one community, the columns of
$\mathbf{S}$ are orthogonal and $\tr (\mathbf{S}^T\mathbf{S})=n$.
The $\delta$-symbol in Equation~\eqref{eq:modularity} can be
expressed as
\begin{equation}\label{eq:deltafunc}
    \delta(c_i,c_j) = \sum_{k=1}^C S_{ik} S_{jk},
\end{equation}
which allows us to express the modularity compactly as
\begin{equation}\label{eq:modularity_matrix_form}
    Q = \frac{1}{2m}\sum_{i,j=1}^n\sum_{k=1}^C B_{ij}S_{ik}S_{jk} =
        \frac{\tr (\mathbf{S}^T \mathbf{B}\mathbf{S})}{2m}.
\end{equation}
This is the quantity that we wish to maximize.

The next step is the `spectral relaxation', where we relax the
discreteness constraints on $\mathbf{S}$, allowing elements of this
matrix to possess real values. We do, however,  constrain the length
of the column vectors by $\mathbf{S}^T\mathbf{S} = \mathbf{M}$,
where $\mathbf{M}$ is a $C\times C$ matrix with the number of nodes
in each community $n_1, n_2,\ldots,n_C$ along the diagonal. In order
to determine the maximum, we take
\begin{equation}\label{eq:relaxed}
    \frac{\partial}{\partial \mathbf{S}}
\left( \frac{1}{2m}\tr [\mathbf{S}^T \mathbf{B} \mathbf{S}] +
\tr[(\mathbf{S}^T \mathbf{S} - \mathbf{M})\tilde{\Lambda}] \right) =
0,
\end{equation}
where $\tilde{\Lambda}$ is a $C\times C$ diagonal matrix of Lagrange
multipliers. The maximum is given by
\begin{equation}\label{eq:matrixsolution}
    \mathbf{B} \mathbf{S} = \mathbf{S} \Lambda,
\end{equation}
where $\Lambda = -2 m \tilde{\Lambda} $ for cosmetical reasons.
Eq.~\eqref{eq:matrixsolution} is a standard matrix eigenvalue
problem. Optimizing in the relaxed representation, we substitute
this solution into Eq.~\eqref{eq:modularity_matrix_form}, and see
that in order to maximize $Q$, we must choose the $C$ largest
eigenvalues of $\mathbf{B}$ and their corresponding eigenvectors.
Since all rows and columns of $\mathbf{B}$ sum to zero by
definition, the vector $(1,1,\ldots,1)^T$ is always an eigenvector
of $\mathbf{B}$ with the eigenvalue $0$. In general the modularity
matrix can have both positive and negative eigenvalues. It is clear
from Eq.~\eqref{eq:modularity_matrix_form} that the eigenvectors
corresponding to negative eigenvalues can never yield a positive
contribution to the modularity. Thus, the number of positive
eigenvalues presents an upper bound on the number of possible
communities.

However, we need to convert our problem back to a discrete one. This
is a non-trivial task. There is no standard way to go from the $n$
continuous entries in each of the $C$ largest eigenvectors of the
modularity matrix and back to discrete $0,1$ values of the community
matrix $\mathbf{S}$. One simple way of circumventing this problem is
to use repeated bisection of the network. This is the procedure that
Newman \cite{newman:06a} recommends. In Newman's scheme, the only
eigenvector utilized is the eigenvector corresponding to the largest
eigenvalue $b_{\rm max}$ of $\mathbf{B}$ (with highest contribution
to $Q$). The $0,1$ vector most parallel to this continuous
eigenvector, is one where the positive elements of the eigenvector
are set to one and the negative elements zero. This is the first
column of the community matrix $\mathbf{S}$. The second column must
contain the remaining elements.

We can increase the modularity iteratively by bisecting the network
into smaller and smaller pieces.  However, this repeated bisection
of the network is problematic. There is no guarantee that that the
best division into three groups can be arrived at by finding by
first determine the best division into two and then dividing one of
those two again. It is straight forward to construct examples where
a sub-optimal division into communities is obtained when using
bisection \cite{newman:06b,reichardt:06a}.

Spectral optimization is not perfect---especially when only the
eigenvector corresponding to $b_{\rm max}$ is
employed\footnote{Newman has proposed a scheme that utilizes two
eigenvectors of the modularity matrix corresponding to the two
highest eigenvalues \cite{newman:06b} that---according to our
experiments---performs slightly better than the single eigenvector
method described above. However, after the application of the
KLN-algorithm described in this section, we found no difference in
the results found by using one or two eigenvectors.}. Therefore,
Newman suggests that it should only be used as a starting point. In
order to improve the modularity, Newman has devised an algorithm
inspired by the classical Kernighan-Lin (KL) scheme
\cite{kernighan:70}. The procedure is as follows: After each
bisection of the network we go through the nodes and find the one
that yields the highest increase in the modularity of the entire
network (or smallest decrease if no increase is possible) if moved
to the other module. This node is now moved to the other module and
becomes inactive. The next step is to go through the remaining $n-1$
nodes and perform the same action. We continue like this until all
nodes have been moved. Finally, we go through all the intermediate
states and pick the one with the highest value of $Q$. This is the
new starting division. We proceed iteratively from this
configuration until no further improvement can be found. Let us call
this optimization the `KLN-algorithm'.

In the spectral optimization, the computational bottleneck is the
calculation of the leading eigenvector(s) of $\mathbf{B}$, which is
non-sparse. Naively, we would expect this to scale like $O(n^3)$.
However, $\mathbf{B}$'s structure allows for a faster calculation.
We can write the product of $\mathbf{B}$ and a vector $\mathbf{v}$
\cite{newman:06a} as
\begin{equation}\label{eq:trick}
    \mathbf{B}\mathbf{v} = \mathbf{A}\mathbf{v} - \frac{\mathbf{k}(\mathbf{k}^T \mathbf{v})}{2m}.
\end{equation}
This way we have a divided the multiplication into (i) sparse matrix
product with the adjacency matrix that takes $O(m+n)$, and (ii) the
inner product $\mathbf{k}^T \mathbf{v}$ that takes $O(n)$. Thus the
entire product $\mathbf{B}\mathbf{v}$ scales like $O(m+n)$. The
total running for a bisection determining the eigenvector(s) is
therefore $O((m+n)n)$ rather than the naive guess of $O(n^3)$. Using
Eq.~\eqref{eq:trick} during the KLN-algorithm reduces the cost of
this step to $O((m+n)n)$ \cite{newman:06a}.

\section{Mean Field Optimization}
Simulated annealing was proposed by Kirkpatrick \emph{et
al}.~\cite{Kirkpatrick83} who noted the conceptual similarity
between global optimization and finding the ground state of a
physical system. Formally, simulated annealing maps the global
optimization problem onto a physical system by identifying the cost
function with the energy function and by considering this system to
be in equilibrium with a heat bath of a given temperature $T$. By
annealing, i.e., slowly lowering the temperature of the heat bath,
the probability of the ground state of the physical system grows
towards unity. This is contingent on whether or not the temperature
can be decreased slowly enough such that the system stays in
equilibrium, i.e., that the probability is Gibbsian
\begin{equation}\label{eq:gibbs}
    P(\mathbf{S}|T) = \frac{1}{Z}\exp\left( - \frac{1}{T} Q(\mathbf{S})\right) =
          \frac{1}{Z}\exp\left( -  \frac{\tr (\mathbf{S}^T \mathbf{B}\mathbf{S})}{2m}\right).
\end{equation}
Here, $Z$ is a constant ensuring proper normalization. Kirkpatrick
et al.\ realized the annealing process by Monte Carlo sampling. The
representation of the constrained modularity optimization problem is
equivalent to a $C$-state Potts model. Gibbs sampling for the Potts
model with the modularity $Q$ as energy function has been
investigated by Reichardt and Bornholdt, see e.g.,
\cite{reichardt:06a}.

Mean field annealing is a deterministic alternative to Monte Carlo
sampling for combinatorial optimization and has been pioneered by
Peterson et al.\ \cite{Peterson87,Peterson89}. Mean field annealing
avoids extensive stochastic simulation and equilibration, which
makes the method particularly well suited for optimization. There is
a close connection between Gibbs sampling and MF annealing. In Gibbs
sampling, every variable is updated by random draw of a Potts state
with a conditional distribution,
\begin{equation}\label{eq:gibbs2}
    P(S_{i1},...,S_{iC}|\mathbf{S}_{\{-i\}},T) = \frac{P(\mathbf{S}|T)}{\sum_{S_{i1},...,S_{iC}}P(\mathbf{S}|T)},
\end{equation}
where the sum runs over the $C$ values of the $i$'th Potts variable
and $\mathbf{S}_{\{-i\}}$ denotes the set of Potts variables
excluding the $i$'th node. As noted by \cite{reichardt:06a},
Eq.~(\ref{eq:gibbs2}) is local in the sense that the part of the
energy function containing variables not connected with the $i$'th
cancels out in the fraction. The mean field approximation is
obtained by computing the conditional mean of the set of variables
coding for the $i$'th Potts variable using Eq.~(\ref{eq:gibbs2}) and
approximating the Potts variables in the conditional probability by
their means \cite{Peterson89}. This leads to a simple
self-consistent set of non-linear
 equations for the means,
\begin{equation}\label{eq:meanfield}
   \mu_{ik}= \frac{\exp(\phi_{ik}/T)}{\sum_{k'=1}^C\exp(\phi_{ik'}/T)}, \ \ \phi_{ik}= \sum_{j} B_{ij}\mu_{jk}.
\end{equation}
For symmetric connectivity matrices with $\sum_j B_{ij} = 0$, the
set of mean field equations has the unique high-temperature solution
$\mu_{ik} = 1/C$. This solution becomes unstable at the mean field
critical temperature, $T_c=b_{\rm max}/C$, determined by the maximal
eigenvalue $b_{\rm max}$ of $\mathbf{B}$.

This mean field algorithm is fast. Each synchronous iteration (see
Section~\ref{sec:experiments} for details on implementation)
requires a multiplication of $\mathbf{B}$ by the mean vector
$\mathbf{\mu}$. As we have seen, this operation can be performed in
$O(m+n)$ time using the trick in Eq.~\eqref{eq:trick}. In these
experiments, we have used a fixed number of iterations of the order
of $O(n)$, which gives us a total of $O((m+n)n)$ similar to the case
of by spectral optimization. (A forthcoming paper discusses the
relationship between Gibbs sampling, mean field methods, and
computational complexity.)

\section{A Simple Network}
\begin{figure}[htbf]
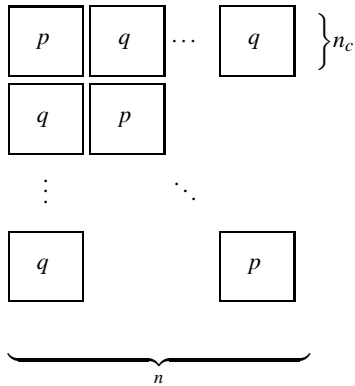

 \begin{displaymath}
 \renewcommand{\arraystretch}{2.5}
 \setlength{\arraycolsep}{.4\arraycolsep}
 \begin{array}{cccccc}
 \sbxs{$p$} & \sbxs{$q$} & \cdots& \sbxs{$q$} & $$\bigg\}n_c$$\\
 \sbxs{$q$} & \sbxs{$p$} &       &  & \\
 \vdots && \ddots & &\\
 \sbxs{$q$} &&& \sbxs{$p$}&\\
 \multicolumn{4}{c}{\underbrace{\hspace{4cm}}_n}&
 \end{array}
 \renewcommand{\arraystretch}{1}
 \end{displaymath}
  \caption{A sketch of the simple network model. The figure displays
  the structure of the adjacency matrix with nodes arranged according to community.
  Inside each community (the blocks)
  along the diagonal, the probability of a link between two nodes is $p$
  and between communities, the probability of a link is $q$.}\label{fig:networksketch}
\end{figure}

We will perform our numerical experiments on a simple model of
networks with communities. This model network consists of $C$
communities with $n_c$ nodes in each, the total network has $n = n_c
C$ nodes. Without loss of generality, we can arrange our nodes
according to their community; a sketch of this type of network is
displayed in Figure~\ref{fig:networksketch}.
Communities are defined as standard random networks, where the
probability of a link between two nodes is given by $p$, with $0< p
\leq 1$. Between the communities the probability of a link between
is given by $0 \leq q < p$. The networks are unweighted and
undirected.

Let us calculate $Q$ for this network in the case where $p=1$ and $q
= 0$. In this case, we can calculate everything exactly. First, we
note that all nodes have the same number of links, and that the
degree of node $i$, $k_i = n_c-1$ (since a node does not link to
itself). Thus the total number of links $m_c$ in each sub-network is
\begin{equation}\label{eq:mc}
    m_c = \frac{1}{2}n_c (n_c -1),
\end{equation}
and since our network consists of $C$ identical communities the
total number of links is $m = C m_c$. We can now write down the
contribution $Q_c$ from each sub-network to the total modularity
\begin{eqnarray}\label{Q_c}
    Q_c &=& \frac{1}{2m}\sum_{ij} (A_{ij} - P_{ij})\delta(c,c)\\
     &=& \frac{1}{2m} \left[n_c (n_c -1) - n_c^2\frac{(n_c -1)^2}{2m}\right].
\end{eqnarray}
If we insert $m$ and use that $Q = C Q_c$, we find
\begin{equation}\label{eq:Qlimit}
    Q = C Q_c = 1 - \frac{1}{C}.
\end{equation}
We see explicitly that when $C\to\infty$ the modularity approaches
unity.

Now, let us examine at the general case. Since our network is
connected at random, we cannot calculate the number of links per
node exactly, but we know that the network is well-behaved (Poisson
link distribution), thus we can calculate the \emph{average} number
of links per node. We see that
\begin{equation}\label{eq:averagedegree}
    k = (n_c - 1)p + n_c(C-1)q,
\end{equation}
which is equal to the number of expected intra-community links plus
the number of expected number of inter-community links. The number
of links in the entire network is therefore given by
\begin{equation}\label{eq:mgeneral}
    m = \frac{1}{2} C n_c k  = \frac{C n_c }{2}[(n_c -1)p +
    n_c(C-1)q].
\end{equation}
We write down $Q$
\begin{eqnarray}
  Q &=& \frac{C}{2m}\left[n_c(n_c-1)p - n_c^2\frac{\{(n_c
- 1)p + n_c(C-1)q\}^2}{2m} \right]\nonumber\\
  &=& \frac{(n_c-1)p}{(n_c-1)p + n_c(C-1) q}-\frac{1}{C}.
\end{eqnarray}
When $n_c \gg 1$ (which is always the case), we have that
\begin{equation}\label{eq:Qfinal}
    Q =\frac{p}{p+q(C-1)}-\frac{1}{C},
\end{equation}
When we write $q$ as some fraction $f$ of $p$, that is $q = fp$,
with $0\leq f \leq 1$, we find
\begin{equation}\label{eq:Qsimple}
    Q(C,f) = \frac{1}{1+(C-1)f}-\frac{1}{C},
\end{equation}
which is independent of $p$. Thus, for this simple network, the only
two relevant parameters are the number of communities and the
density of the inter-community links relative to the intra-community
strength. We can also see that our result from Eq.~\eqref{eq:Qlimit}
is valid even in the case $p < 1$, as long as the communities are
connected and $q=0$.
\begin{figure}
  \includegraphics[width=\hsize]{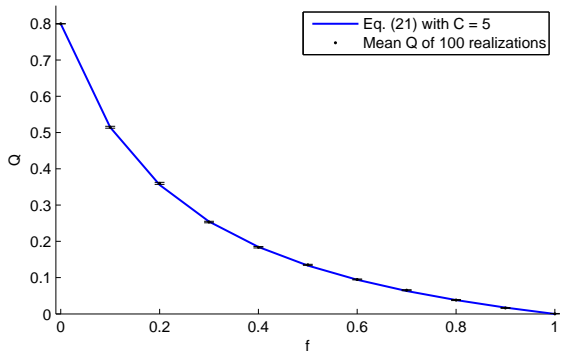}\\
  \caption{Equation~\eqref{eq:Qsimple} and $Q_\text{design}$. This figure displays
  $Q$ as a function of $f$ (the relative probability of a link
  between communities), with $C=5$ for the simple network
  defined in Figure~\ref{fig:networksketch}. The blue line is given
  by Eq.~\eqref{eq:Qsimple} and the black dots with error-bars are mean
  values of $Q_{\text{design}}$ in realizations of the simple network
  with $p = 1/10$ and $n=500$; each data-point is the mean of $100$
  realizations. The error bars are calculated as the standard deviation
  divided by square root of the number of runs.}\label{fig:curve}
\end{figure}

If we design an adjacency matrix according to
Figure~\ref{fig:networksketch}, we can calculate the value
$Q_{\text{design}} = \tr
(\mathbf{S}_{\text{d}}^T\mathbf{B}\mathbf{S}_{\text{d}})/(2m)$,
where $\mathbf{S}_\text{d}$ is a community-matrix that reflects the
designed communities. Values of $Q_{\rm design}$ should correspond
to Eq.~\eqref{eq:Qsimple}. We see in Figure~\ref{fig:curve} that
this expectation is indeed fulfilled. The blue curve is $Q$ as a
function of $f$ with $C=5$. The black dots with error-bars are mean
values of $Q_{\text{design}}$ in realizations of the simple network
with $p = 1/10$ and $n=500$; each data-point is the mean of $100$
realizations and the error bars are calculated as the standard
deviation divided by square root of the number of runs. The
correspondence between prediction and experiment is quite
compelling.

We should note, however, that the value of $Q_{\rm design}$ may be
lower than the actual modularity found for the network by a good
algorithm: We can imagine that fluctuations of the inter-community
links could result in configurations that would yield higher values
of $Q$---especially for high values of $f$. We can quantify this
quite precisely. Reichardt and Bornholdt \cite{reichardt:06a} have
shown that demonstrated that random networks can display
significantly larger values of $Q$ due to fluctuations; when $f =
1$, our simple network is precisely a random network (see also
related work by Guimer\`a \emph{et al.} \cite{guimera:04a}). In the
case of the network we are experimenting on, ($n=500$, $p = 1/10$),
they predict $Q \approx 0.13$.

Thus, we expect that the curve for $Q(f,C)$ with fixed $C$ will be
deviate from the $Q_\text{design}$ displayed in
Figure~\ref{fig:curve}; especially for values of $f$ that are close
to unity. The line will decrease monotonically from $Q(0,C)=1-1/C$
towards $Q(1,C) = 0.11$ with the difference becoming maximal as
$f\to 1$.

\section{Numerical Experiments}\label{sec:experiments}
We know that the running time of mean field method scales like that
of the spectral solution. In order to compare the precision of the
mean field solutions to the solutions stemming from spectral
optimization, we have created a number of test networks with
adjacency matrices designed according to
Figure~\ref{fig:networksketch}. We have created $100$ test networks
using parameters $n_c = 100$, $C = 5$, $p=0.1$ and $f \in [0,1]$.
Varying $f$ over this interval allows us to interpolate between a
model with $C$ disjunct communities and a random network with no
community structure.

We applied the following three algorithms to our test networks
\begin{enumerate}
  \item Spectral optimization,
  \item Spectral optimization and the KLN-algorithm, and
  \item Mean field optimization.
\end{enumerate}
Spectral optimization and the KLN-algorithm were implemented as
prescribed in \cite{newman:06a}. The $n  C$ non-linear mean field
annealing equations were solved approximately using a $D=300$-step
{\it annealing schedule} linear in $\beta = 1/T$ starting at
$\beta_c $ and ending in $3\beta_c$ at which temperature the
majority of the mean field variables are saturated. The mean field
critical temperature $T_c = b_{\rm max}/C$ is determined for each
connectivity matrix. The synchronous update scheme defined as
parallel update of all means at each of the $D$ temperatures
\begin{eqnarray}
  \mu_{ik}^{(d+1)}&=& \frac{\exp(\phi_{ik}^{(d)}/T)}{\sum_{k'=1}^C\exp(\phi_{ik'}^{(d)}/T)}\nonumber\\
 \phi_{ik}^{(d)}&=& \sum_{j} B_{ij}\mu_{jk}^{(d)}\label{eq:update}
\end{eqnarray}
can grow unstable at low temperatures. A slightly more effective and
stable update scheme is obtained by selecting random fractions $\rho
< 1$ of the means for update in $1/\rho$ steps at each temperature.
We use $\rho=0.2$ in the experiments reported below. A final $T=0$
iteration, equivalent to making a decision on the node community
assignment, completes the procedure. We \emph{do not} assume that
actual the number of communities $C < C_{\rm max}$ is known in
advance. In these experiments we use $C_{max} = 8$. This number is
determined after convergence by counting the number of non-empty
communities

\begin{figure}
  \includegraphics[width=\hsize]{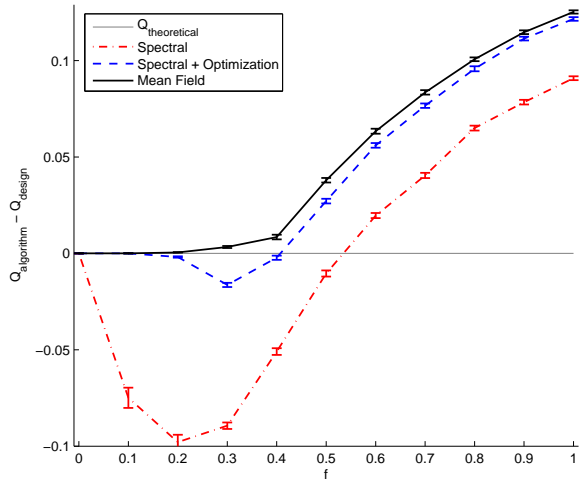}\\
  \caption{Comparing spectral methods with the mean field solution.
  The networks were created according to the simple model, using
  parameters $n_c = 100$, $C = 5$, $p=0.1$ and $f \in [0,1]$.
  All data points display the point-wise differences between
  the value of $Q_\text{algorithm}$ found by the algorithm in question
  and $Q_\text{\rm design}$. The error-bars are calculated as in
  Figure~\ref{fig:curve}. The dash-dotted red line shows the results
  for the spectral method. The dashed blue line shows the results for
  the spectral optimization followed by KLN post-processing. The solid
  black curve shows the results for the mean field optimization. The grey,
  horizontal line corresponds to the theoretical prediction (Eq.~\eqref{eq:Qsimple})
  for the designed communities.
  }
  \label{fig:compare}
\end{figure}

The results of the numerical runs are displayed in
Figure~\ref{fig:compare}. This figure shows the point-wise
differences between the value of $Q_\text{algorithm}$ found by the
algorithm in question and $Q_\text{\rm design}$ plotted as a
function of the inter-community noise $f$.
The line of $Q_\text{algorithm} - Q_{\rm design} = 0$ thus
corresponds to the curve plotted in Figure~\ref{fig:curve}. We see
from Figure~\ref{fig:compare} that the mean field approach uniformly
out-performs both spectral optimization and spectral optimization
with KLN post-processing. We also ran a Gibbs sampler
\cite{reichardt:06a} for with a computational complexity equivalent
to the mean field approach. This lead to communities with $Q$
slightly lower than the mean field results, but still better than
spectral optimization with KLN post-processing.

We note that the obtained  $Q_{\rm algorithm}$ for a random network
($f = 1$) is consistent with the prediction made by Reichardt and
Bornholdt \cite{reichardt:06a}. We also see that the optimization
algorithms can exploit random connections to find higher values of
$Q_{\rm algorithm}$ than expected for the designed communities
$Q_{\rm design}$. In the case of the mean field algorithm this
effect is visible for values of $f$ as low as $0.2$.

\begin{figure}
  \includegraphics[width=\hsize]{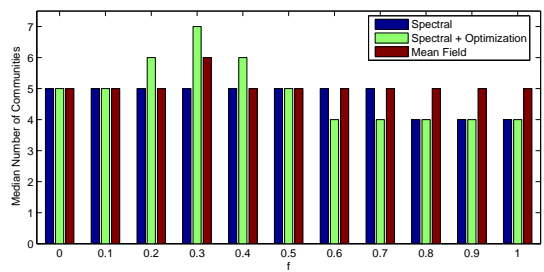}\\
  \caption{The median number of communities found
  by the various algorithms. The panel shows the median number of
  communities as a function of the relative fraction of inter-community links
  $f$. All optimization schemes consistently pick four or five
  communities for the highest values of $f$. This finding is consistent
  with theoretical and experimental results by Reichardt and Bornholdt
  \cite{reichardt:06a}
  }
  \label{fig:communities}
\end{figure}

Figure~\ref{fig:communities} shows the median number of communities
found by the various algorithms as a function of $f$.
It is evident from Figs.~\ref{fig:compare} and~\ref{fig:communities}
that---for this particular set of parameters---the problem of
detecting the designed community structure is especially difficult
around $f = 0.3$. Spectral clustering with and without the KLN
algorithm find values $Q_{\rm algorithm}$ that are significantly
lower than $Q_{\rm desigm}$. The mean field algorithm manages to
find a value of $Q_{\rm algorithm}$ that is higher than the designed
$Q$ but does so by creating extra communities. As $f \to 1$ it
becomes more and more difficult to recover the designed number of
communities.

\section{Conclusions}
We have introduced a deterministic mean field annealing approach to
optimization of modularity $Q$. We have evaluated the performance of
the new algorithm within a family of networks with variable levels
of inter-community links, $f$. Even with a rather costly
post-processing approach, the spectral clustering approach suggested
by Newman is consistently out-performed by the mean field approach
for higher noise levels. Spectral clustering without the KLN
post-processing finds much lower values of $Q$ for all $f > 0$.

Speed is not the only benefit of the mean field approach. Another
advantage is that the implementation of mean field annealing is
rather simple and similar to Gibbs sampling. This method also avoids
the inherent problems of repeated bisection. The deterministic
annealing scheme is directed towards locating optimal configurations
without wasting time at careful thermal equilibration at higher
temperatures. As we have noted above, the modularity measure $Q$ may
need modification in specific non-generic networks. In that case, we
note that the mean field method is quite general and can be
generalized to many other measures.

\end{document}